\documentclass{emulateapj}

\slugcomment{Accepted for Publication in The Astrophysical Journal}

\shorttitle{A Giant Ly$\alpha$ Emitter at $z=6.595$}
\shortauthors{Ouchi et al.}

\begin{document}

\title{Discovery of a Giant Ly$\alpha$ Emitter Near the Reionization Epoch 
      \altaffilmark{1,2}}

\author{
Masami Ouchi        \altaffilmark{3,4},
Yoshiaki Ono        \altaffilmark{5},
Eiichi Egami        \altaffilmark{6},
Tomoki Saito        \altaffilmark{7},
Masamune Oguri      \altaffilmark{8},\\
Patrick J. McCarthy    \altaffilmark{3},
Duncan Farrah       \altaffilmark{9,10},
Nobunari Kashikawa  \altaffilmark{11},
Ivelina Momcheva    \altaffilmark{3},\\
Kazuhiro Shimasaku  \altaffilmark{5},
Kouichiro Nakanishi \altaffilmark{12},
Hisanori Furusawa   \altaffilmark{13},
Masayuki Akiyama    \altaffilmark{14},\\
James S. Dunlop     \altaffilmark{15,16},
Angela M. J. Mortier \altaffilmark{16},
Sadanori Okamura    \altaffilmark{5},
Masao Hayashi       \altaffilmark{5},\\
Michele Cirasuolo   \altaffilmark{16},
Alan Dressler       \altaffilmark{3},
Masanori Iye        \altaffilmark{11},
Matt. J. Jarvis     \altaffilmark{17},\\
Tadayuki Kodama     \altaffilmark{11},
Crystal L. Martin   \altaffilmark{18},
Ross J. McLure      \altaffilmark{16},\\
Kouji Ohta          \altaffilmark{19},
Toru Yamada         \altaffilmark{14},
Michitoshi Yoshida  \altaffilmark{20}
}

\altaffiltext{1}{Based in part on data collected at Subaru Telescope, which is operated by the National Astronomical Observatory of Japan.}
\altaffiltext{2}{Some of the data presented herein were 
                obtained at the W.M. Keck Observatory, 
                which is operated as a scientific partnership 
                among the California Institute of Technology, 
                the University of California and the National 
                Aeronautics and Space Administration. 
                The Observatory was made possible by the generous 
                financial support of the W.M. Keck Foundation.}
\altaffiltext{3}{Observatories of the Carnegie Institution of Washington,
        813 Santa Barbara St., Pasadena, CA 91101}
\altaffiltext{4}{Carnegie Fellow; ouchi \_at\_ ociw.edu}
\altaffiltext{5}{Department of Astronomy, School of Science, University of Tokyo, Tokyo 113-0033, Japan}
\altaffiltext{6}{Department of Astronomy, University of Arizona, 933 N Cherry Avenue, Rm. N204, Tucson, AZ 85721-0065}
\altaffiltext{7}{Research Center for Space and Cosmic Evolution, Ehime University, 2-5 Bunkyo-cho, Matsuyama 790-8577, Japan}
\altaffiltext{8}{Kavli Institute for Particle Astrophysics and Cosmology, Stanford University, 2575 Sand Hill Road, Menlo Park, CA 94025}
\altaffiltext{9}{Department of Astronomy, Cornell University, Ithaca, NY 14853}
\altaffiltext{10}{Astronomy Centre, University of Sussex, Falmer, Brighton, UK}
\altaffiltext{11}{Optical and Infrared Astronomy Division, National Astronomical Observatory, Mitaka, Tokyo 181-8588, Japan}
\altaffiltext{12}{Nobeyama Radio Observatory, Minamimaki, Minamisaku, Nagano 384-1305, Japan}
\altaffiltext{13}{Subaru Telescope, National Astronomical Observatory of Japan, 650 North A'ohoku Place, Hilo, HI 96720}
\altaffiltext{14}{Astronomical Institute, Graduate School of Science, Tohoku University, Aramaki, Aoba, Sendai 980-8578, Japan}
\altaffiltext{15}{Department of Physics and Astronomy, University of British Columbia, 6224 Agricultural Road, Vancouver V6T 1Z1, Canada}
\altaffiltext{16}{SUPA Institute for Astronomy, University of Edinburgh, Royal Observatory, Edinburgh EH9 3HJ, UK}
\altaffiltext{17}{Centre for Astrophysics, Science \& Technology Research Institute, University of Hertfordshire, Hatfield AL10 9AB, UK}
\altaffiltext{18}{Department of Physics, University of California, Santa Barbara, CA 93106}
\altaffiltext{19}{Department of Astronomy, Kyoto University, Kyoto 606-8502, Japan}
\altaffiltext{20}{Okayama Astrophysical Observatory, National Astronomical Observatory, Kamogata, Okayama 719-0232, Japan}

\begin{abstract}
We report the discovery of a giant Ly$\alpha$ emitter (LAE)
with a Spitzer/IRAC counterpart 
near the reionization epoch at $z=6.595$.
The giant LAE is found from the extensive 1 deg$^2$ Subaru 
narrow-band survey for $z=6.6$ LAEs in the 
Subaru/XMM-Newton Deep Survey (SXDS) field, and subsequently 
identified by deep spectroscopy of Keck/DEIMOS and Magellan/IMACS.
Among our 207 LAE candidates, this LAE is not only the brightest 
narrow-band object with $L({\rm Ly}\alpha) = 3.9 \pm 0.2 \times 10^{43}$ erg s$^{-1}$ 
in our survey volume of $10^6$ Mpc$^3$, but also 
a spatially extended Ly$\alpha$ nebula with the largest isophotal area
whose major axis is at least $\simeq 3''$. 
This object is more likely to be a large Ly$\alpha$ nebula 
with a size of $\gtrsim 17$-kpc than 
to be a strongly-lensed galaxy by a foreground object.
Our Keck spectrum with medium-high spectral and spatial resolutions 
suggests that the velocity width is $v_{\rm FWHM}=251\pm 21$ km s$^{-1}$,
and that the line-center velocity changes by $\simeq 60$ km s$^{-1}$ 
in a $10$-kpc range.
The stellar mass and star-formation rate are 
estimated to be $0.9-5.0\times 10^{10} M_\odot$ and $>34$ $M_\odot$yr$^{-1}$,
respectively, from the combination of deep optical to infrared images of Subaru, 
UKIDSS-Ultra Deep Survey, and Spitzer/IRAC.
Although the nature of this object is not yet clearly
understood, this could be an important object for studying
cooling clouds accreting onto a massive halo,
or forming-massive galaxies with significant outflows contributing to
cosmic reionization and metal enrichment of inter-galactic medium.
\end{abstract}


\section{Introduction}
\label{sec:introduction}

Identifying the first stage of galaxy formation 
is one of the ultimate goals in astronomy today.
Theoretical models predict that primordial gas accretes
onto the center of halos via gravitational cooling
with subsequent star-formation activity
\citep{fardal2001,yang2006}.
These primordial galaxies make spatially extended Ly$\alpha$ nebulae
caused by hydrogen cooling, and it is suggested that 
high-$z$ extended Ly$\alpha$ sources, or Ly$\alpha$ blobs, 
are candidates for primordial galaxies
(e.g. \citealt{matsuda2004,saito2006,nilsson2006,smith2007}).
Ly$\alpha$ blobs are found mostly at $z\simeq 2-3$, and have
angular extents of $\simeq 5-16$ arcsec with total Ly$\alpha$
luminosities ranging from $\simeq 6\times 10^{42}$ to $10^{44}$ erg s$^{-1}$
\citep{matsuda2004}.
The most prominent Ly$\alpha$ nebulae known to date are blobs 1 and 2 
found by \citet{steidel2000}, which extend over $\gtrsim 100$ kpc
with $L(Ly\alpha)\simeq 10^{44}$ erg s$^{-1}$.
Although Ly$\alpha$ blobs are candidates for galaxies with gas
inflow of cooling accretion, 
it is also suggested that Ly$\alpha$ blobs can be produced 
by intensive starbursts associated with significant outflows
(e.g., \citealt{taniguchi2000,wilman2005}), 
by a hidden AGN (e.g., \citealt{haiman2001}), 
or by both of them (e.g., \citealt{dey2005}; Yang et al. in preparation).
\begin{figure}
\epsscale{1.1}
\plotone{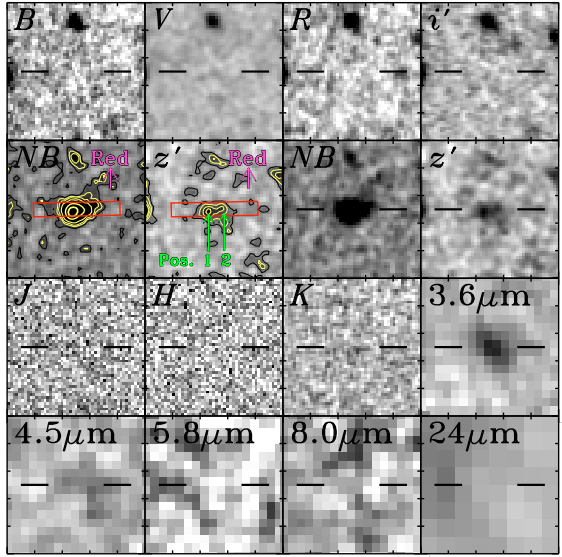}
\caption{
Optical to infrared images of Himiko.
North is up and east is to the left.
We display $10''\times10''$ images at $BVRi'z'$ and $NB$ ($NB921$) bands 
from Subaru/SXDS, at $JHK$ bands from UKIDSS-UDS DR3, 
and $3.6-24\mu$m bands from Spitzer/SpUDS.
We show intensity contours in $NB921$ and $z'$ images.
The black contours denote 1 $\sigma$ level of sky fluctuation.
The yellow contours represent (2, 3, 5, 10, 15) and (2,3,4) $\sigma$ levels of 
sky fluctuations in $NB921$ and $z'$ images, respectively.
We also plot
the position of the DEIMOS slit by the red box. 
The dispersion direction towards red spectrum 
is shown by the magenta arrow.
The green arrows point to
the $position$ 1 and 2 that are probable peaks 
in the $z'$ image.
\label{fig:z7blob_images}}
\end{figure}
In fact, the infrared-submm and X-ray observations
indicate that \citeauthor{steidel2000}'s blobs 1 and 2
would be powered by a heavily obscured starburst 
(\citealt{geach2007,matsuda2007}; see also \citealt{chapman2004}) 
and an AGN \citep{basu-zych2004}, respectively.
\citet{matsuda2006} claim
that all of their spectroscopically-identified Ly$\alpha$ blobs
are likely to be the sites of massive galaxy formation 
because of their large line widths of $v_{\rm FHWM}\gtrsim 500$ km s$^{-1}$.
It is also well known that such bright large Ly$\alpha$ nebulae are associated with
radio-loud AGN (e.g. \citealt{mccarthy1987,vanojik1997,reuland2003,barrio2008,smith2009}) or
radio-quiet quasars (\citealt{weidinger2005,hennawi2008}).
In this way, extended Ly$\alpha$ nebulae shed light not only on
primordial galaxies but also on massive-galaxy formation and AGN activity.

\begin{figure}
\epsscale{1.1}
\plotone{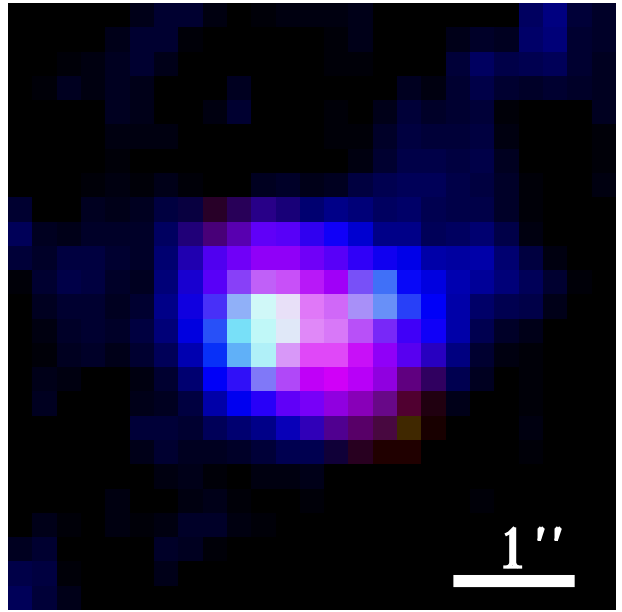}
\caption{
Composite pseudo-color image of Himiko.
The RGB colors are assigned to
$3.6\mu$m , $z'$, and $NB921$ images, respectively.
North is up and east is to the left.
The image size is $5''\times5''$. 
The white bar at the bottom right represents the length of 
one arcsecond. The brightest peak with a bluish white color
corresponds to $position$ 1. The $position$ 2 is located
$1.1$ arcsec west of the $position$ 1.
\label{fig:z7blob_composite}}
\end{figure}

So far, extended Ly$\alpha$ nebulae have been found only at $z=2-5$ 
with the majority at $z\simeq 2-3$ (e.g., \citealt{saito2008}).
Due to this current observational limit, it is difficult 
to identify primordial galaxies as well as 
to study the early stage of massive galaxy formation.
At $z\simeq 2-3$, the mean metallicity of the inter-stellar medium is 
already as high as $Z=0.1 Z_\odot$ (\citealt{sadat2001}; 
see also \citealt{pettini1997}). It is predicted that
the fraction of primordial galaxies to metal enriched galaxies 
would be quite low at $z\simeq 2-3$ (\citealt{scannapieco2003}),
and that the fraction of population III to population II 
star-formation rate (SFR)
rises with increasing redshift \citep{trac2007}.
On the other hand, the importance of the early stage of massive-galaxy formation
has been recognized by the downsizing behavior of stellar-mass
assembly \citep{cowie1996}.
A massive population at $z=2-3$ selected from a distant-red galaxy sample 
is old, $2-3$ Gyr, and their mean formation redshift 
is estimated to be $z\gtrsim 5$ (\citealt{labbe2005};
see also \citealt{kriek2006}).
It is implied that galaxies at the massive end would have a very high
specific star-formation rate (SSFR) at $z\gtrsim 4$ \citep{drory2008}, and
that a major active star-formation in massive galaxies probably takes place
at $z\gtrsim 4$.
Thus, it is important to study extended Ly$\alpha$ nebulae
at a redshift greater than the current-observational limit,
especially at the reionization epoch of $z\simeq 6-11$ \citep{fan2006,komatsu2008}. 
This epoch is also the today's observational limit of galaxy studies
\citep{iye2006,stark2007,ota2008,bouwens2008}. Moreover, such bright Ly$\alpha$ sources
can be a good laboratory for understanding
reionization (e.g. \citealt{kashikawa2006,dijkstra2007b,mcquinn2007,kobayashi2007}) 
and metal enrichment (e.g. \citealt{martin2002,bouche2007}) of inter-galactic medium (IGM).

In this paper, we report our discovery of an extended
Ly$\alpha$ nebula, which we named Himiko
\footnote{
Himiko is a name of legendary queen in ancient Japan.
}
, near the reionization epoch
at $z=6.595$.
We describe the photometric identification and spectroscopic confirmation of this object 
in \S \ref{sec:discovery},
and present detailed properties such as kinematics and stellar population 
in \S \ref{sec:giant_LAE}.
We discuss the nature of this object and prospects of future observations
in \S \ref{sec:discussion}.
Throughout this paper, magnitudes are in the AB system. 
We adopt 
$(h,\Omega_m,\Omega_\Lambda,n_s,\sigma_8)=(0.7,0.3,0.7,1.0,0.8)$.

\begin{figure}
\epsscale{0.88}
\plotone{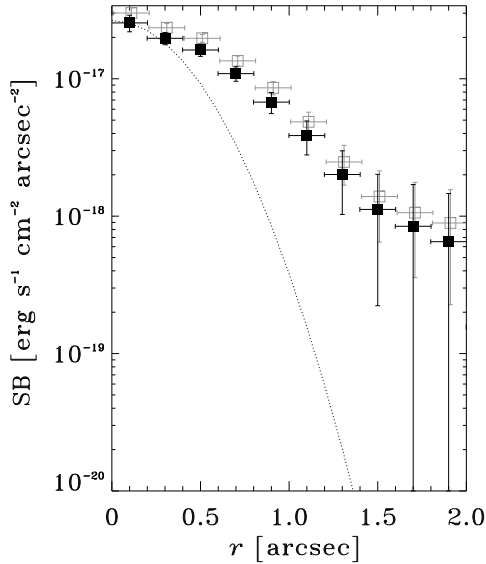}
\caption{
Surface brightness (SB) profiles of our object.
Open squares show a SB profile in $NB921$ image,
and filled circles denote that of a continuum-subtracted
(Ly$\alpha$) image.
The dotted line is the Gaussian profile representing
the point-spread function of our image with an $FWHM=0''.8$.
\label{fig:radial_profile_z7blob}}
\end{figure}

\vspace{2cm}

\section{Discovery}
\label{sec:discovery}

\subsection{Photometric Identification}
\label{sec:photometric_identification}

We have identified a candidate 
very bright spatially-extended Ly$\alpha$ emitter (LAE) 
at $z\simeq 6.6$ in the course of our deep 
and wide-field narrow-band imaging program in the 
Subaru/XMM-Newton Deep Survey (SXDS) field 
\citep{ouchi2008}. In 2005-2007, we took narrow-band images in 
the $NB921$ filter with a central wavelength of 
$\lambda_c=9196$\AA\ and a FWHM of 132\AA\ \citep{hayashino2003}
using Subaru/Suprime-Cam \citep{miyazaki2002}. 
The 1 deg$^2$ field is covered by 5 pointings of Suprime-Cam
with a total on-source integration of 45.1 hours. 
These data are reduced with SDFRED \citep{yagi2002,ouchi2004},
and aligned with optical broad-band images of SXDS \citep{furusawa2008}.
The FWHM of the seeing size in the aligned images is $\simeq 0''.8$.
The $3\sigma$ limiting magnitude in $NB921$ is 
$26.2-26.4$ mag in a $2''.0$-diameter aperture.
Combining the deep optical broad-band images of SXDS,
we have selected candidates of $z\simeq 6.56\pm0.05$ LAEs that
satisfy our photometric criteria of 
the narrow-band excess ($z'-NB921>1.0$),
no detection of blue continuum flux ($B>B_{2\sigma}$ and $V>V_{2\sigma}$),
and
the existence of Gunn-Peterson trough 
($[z'<z'_{3\sigma}$ \& $i'-z'>1.3]$ or $[z'\ge z'_{3\sigma}]$).
The $B_{2\sigma}$ and $V_{2\sigma}$ are defined as $2\sigma$ limiting magnitudes of
$B$ and $V$ bands, respectively ($B_{2\sigma}=28.7$ and $V_{2\sigma}=28.2$), 
while $z'_{3\sigma}$ is the $3\sigma$ detection limit ($z'_{3\sigma}=26.5$). 
We have obtained a photometric sample of 207 LAEs at $z\simeq 6.6$ down to $NB921=26.0$
in a comoving survey volume of $8\times 10^5$ Mpc$^3$.
The sky distribution of our LAEs
show a rectangular area ($8'\times 20'$)
with a number density of LAEs higher than the average
by a factor of 2.
In this high-density region,
we find the object, Himiko, 
that has the brightest $NB921$ magnitude and the largest isophotal area 
among the 207 LAE candidates.
\begin{figure}
\epsscale{1.2}
\plotone{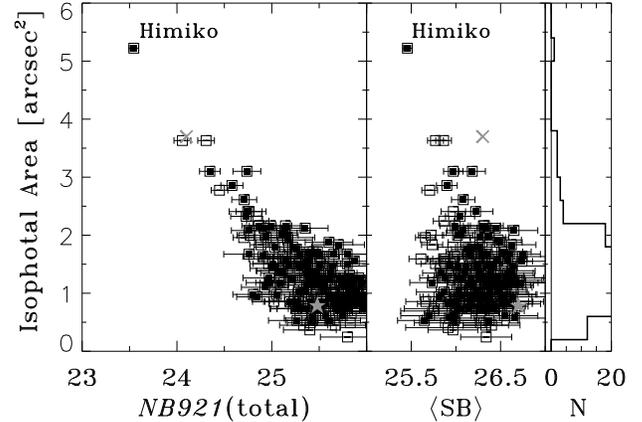}
\caption{
Isophotal area of our $z=6.6$ LAE candidates as a function
of total $NB921$ magnitude (left panel) and
average $NB921$ surface brightness, $\left< SB \right>$ (middle panel).
$\left< SB \right>$ is expressed in units of mag arcsec$^{-2}$.
The square with a label of Himiko is our giant LAE.
The open squares with a filled square represent
$z=6.6$ LAE candidates showing a possibly extended profile
with a FWHM of $>1''.2$ in the $NB921$ image (a PSF FWHM of
$0''.8$), while the simple open squares indicate 
the other (FWHM$\simeq 0''.8-1''.2$) candidates.
The measurements of FWHM include large uncertainties
for faint sources with $NB921\simeq 25-26$, and
become unreliable in this faint magnitude regime.
The gray star and cross marks denote our simulated blob 1 and 28,
respectively (see \S \ref{sec:comparisons}).
The right panel plots the isophotal-area distribution of 
the LAE candidates with the thick line. The thin line presents $N=0$.
\label{fig:mag_isoarea_NB921LAE}}
\end{figure}
We present snapshot images in Figure \ref{fig:z7blob_images}
and a close-up color composite image in Figure \ref{fig:z7blob_composite}.
The total magnitude of Himiko is $NB921=23.55$, which is brighter than 
the second brightest candidate ($NB921=24.06$) by $0.5$ magnitude.
This object is significantly extended,
in contrast to the compact point-like profiles of the other LAEs.
In Figure \ref{fig:radial_profile_z7blob}, 
we plot surface brightness profiles of this object
in $NB921$ image and a continuum-subtracted Ly$\alpha$ image
obtained from $NB921$ and $z'$ data with the assumption of 
$z=6.6$ LAE spectrum shape \citep{shimasaku2006}.
Figure \ref{fig:radial_profile_z7blob} shows that 
the profiles of this object is clearly more extended
than that of point-spread function (PSF) of our image ($FWHM=0''.8$).
The outskirts of the profile reach a radius up to 
$\simeq 1''.5$ and a possibly to $\sim 2''$.
If we define the isophotal area, $A_{\rm iso}$, as pixels with values above 
the $2 \sigma$ sky fluctuation ($26.8$ mag arcsec$^{-2}$ in $NB921$),
the isophotal area of Himiko is $A_{\rm iso}=5.22$ arcsec$^2$ 
in the $NB921$ image.
Figure \ref{fig:mag_isoarea_NB921LAE} presents the isophotal area of
our $z=6.6$ LAE candidates as a function of total $NB921$ magnitude
and average $NB921$ surface brightness. 
The average $NB921$ surface brightness, $\left< SB \right>$,
is the value of an isophotal flux divided by the isophotal area,
where the isophotal flux is the one summed over the isophotal area.
We mark a possibly extended
(FWHM$>1''.2$) sources with a filled squares
to distinguish between bright point-like and faint extended 
sources with a comparable isophotal area.
The right panel of Figure \ref{fig:mag_isoarea_NB921LAE}
plots an isophotal-area distribution with a histogram,
which visualizes the extended nature of our object
among all of the LAE candidates.
Figure \ref{fig:mag_isoarea_NB921LAE} indicates
that there are no LAEs similar to Himiko.
We confirm that the brightest source from the previous 
0.2 deg$^2$ Subaru Deep Field (SDF) survey for $z\sim 6.5$ LAEs 
is only as bright as our second brightest candidate 
with no significant spatial extent 
\citep{taniguchi2005,kashikawa2006}, and that
our object is distinguished from all the other $z\sim 6.5$ 
LAEs found in the previous studies.
By the comparisons with $z\sim 6-7$ LAEs found in previous studies,
we notice that the size of this object is outstanding.
The major axes of the isophotal area in $NB921$ and $z'$ bands
are $\simeq 3''.1$ and $\simeq 2''.0$, respectively.
Additionally, the $NB921$ ($z'$) image shows potential diffuse components 
which continuously extend by $\sim 1''$ ($\sim 0''.3$) 
around the isophotal area
with a surface brightness above $1 \sigma$ sky fluctuation
(Figures \ref{fig:z7blob_images} and \ref{fig:z7blob_composite}).
Thus, the size of our object is probably $\gtrsim 3''.1$ and 
$\gtrsim 2''.0$ in $NB921$ and $z'$ bands, respectively.
Given the fact that this LAE has the unusual brightness and size, 
we refer to this object as the giant LAE.

Interestingly, this object is detected at the $4\sigma$ level 
in the medium deep $3.6\mu$m image from the Spitzer legacy survey of 
the Ultra Deep Survey field (SpUDS; PI:. J. Dunlop;
Figure \ref{fig:z7blob_images}),
while we find only marginal detections ($\sim 2-3\sigma$)
\footnote{
We estimate the $2\sigma$ limits of total magnitudes
in the vicinity of this object to be $J=24.3$, $H=24.0$, and $K=23.8$.
}
in the near-infrared (NIR) images from
the UKIRT Infrared Deep Sky Survey Third Data Release 
(UKIDSS-DR3: \citealt{lawrence2007}).
We align Spitzer/SpUDS and UKIDSS-DR3 images
with the SXDS optical images, referring 
a number of stellar objects in the field.
The relative astrometric errors are estimated to be
$\simeq 0''.04$, $\simeq 0''.11$, and $\simeq 0''.35$
in rms, for optical-NIR, Spitzer/IRAC($3.6-8.0\mu$m), and MIPS($24\mu$m) images,
respectively.
We summarize total magnitudes/fluxes and $2''$-diameter aperture
magnitudes of Himiko in Table \ref{tab:photometry}.
We define the total magnitude with MAG\_AUTO of SExtractor \citep{bertin1996}
in the optical and NIR bands. The total magnitudes of Spitzer/IRAC and MIPS 
bands are obtained from a $3''$-diameter aperture and an aperture
correction given in \citet{yan2005} and the MIPS web page
\footnote{
http://ssc.spitzer.caltech.edu/mips/apercorr/
}, 
respectively.
Note that our object is detected in the $3.6\mu$m band, but
not in the $4.5\mu$m band. This is probably due to the
higher noise level in the $4.5\mu$m band, as we expect the object
to have a fairly flat spectrum at these wavelengths. 
Our measured $3.6\mu$m magnitude of $24.02$ ($4\sigma$) would 
result in 
a $<3\sigma$ detection at $4.5\mu$m for a flat
spectrum (constant AB magnitude), consistent with 
our tentative $1-2\sigma$ detection and the large error
with $24.62\pm 0.73$
\footnote{
See \S \ref{sec:agn} for a possible inclusion of emission lines in the IRAC bands.
}.

\begin{deluxetable}{ccc}
\tabletypesize{\scriptsize}
\tablecaption{Photometry of Himiko
\label{tab:photometry}}
\tablewidth{0pt}
\tablehead{
\colhead{Band} &
\colhead{Mag($2''$)} &
\colhead{Mag/Flux(Total)}\\
\colhead{} &
\colhead{(1)} &
\colhead{(2)}
}
\startdata
$f(0.5-2{\rm keV})$\tablenotemark{2} & \nodata & $<6\times 10^{-16}$\\
$B$ & $>28.7$ & $>27.9$ \\
$V$ & $>28.2$ & $>27.4$ \\
$R$ & $>28.1$ & $>27.3$ \\
$i'$ & $>28.0$ & $>27.2$ \\
$z'$\tablenotemark{3} & $25.86\pm 0.20$ & $25.45\pm 0.27$ \\  
$NB921$\tablenotemark{3} & $23.91\pm 0.04$ & $23.55\pm0.05$ \\
$m_{0.95}$ & $25.74 \pm 0.64$ & $25.18\pm 0.73$ \\ 
$J$\tablenotemark{4} & $24.95\pm 0.53$ & $24.01\pm 0.43$ \\
$H$\tablenotemark{4} & $>24.7$ & $>24.0$ \\
$K$\tablenotemark{4} & $24.42\pm 0.50$ & $>23.8$ \\ 
$m(3.6\mu{\rm m})$ & \nodata & $24.02\pm 0.27$ \\
$m(4.5\mu{\rm m})$ & \nodata & $>23.9 $ \\
$m(5.8\mu{\rm m})$ & \nodata & $>22.0$ \\
$m(8.0\mu{\rm m})$ & \nodata & $>21.8$ \\
$m(24\mu{\rm m})$ & \nodata & $>19.8$ \\
$S(850\mu m)$ & \nodata & $<12$mJy \\
$f(1.4{\rm GHz})$ & \nodata & $<100\mu$Jy\\
\enddata

\tablecomments{
Col.(1): The $2''$-diameter aperture magnitude. Col.(2): The total magnitude or flux.
In these two columns, the upper limits are $2\sigma$ and $3\sigma$ magnitudes 
in $B-K$ and $3.6-24\mu$m bands, respectively.
}
\tablenotetext{2}{In units of erg cm$^{-2}$ s$^{-1}$.}
\tablenotetext{3}{Isophotal magnitudes are $25.67\pm 0.21$ and $23.66\pm0.04$ in $z'$ and $NB921$ bands,
respectively.}
\tablenotetext{4}{Magnitudes in $J$ and $K$ bands are slightly over the $2\sigma$ level. However, neither of
them are detected beyond the $3\sigma$ level.}
\end{deluxetable}

\begin{figure}
\epsscale{1.2}
\plotone{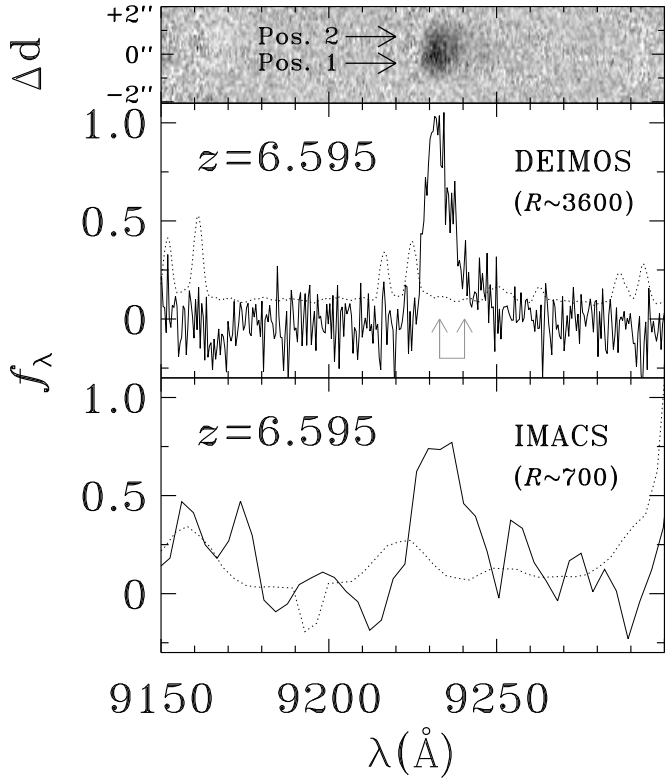}
\caption{
Spectra of the giant LAE, Himiko. The top panel shows the
two-dimensional spectrum obtained from DEIMOS observations. 
The horizontal arrows point to the $position$ 1 and 2.
The middle and bottom
panels present the spectra taken with DEIMOS and IMACS, respectively.
We show spectra of the giant LAE (solid line) and the background sky (dotted line).
The units of the vertical axis are
in $10^{-17}$ erg s$^{-1}$ cm$^{-2}$ \AA$^{-1}$ in the bottom panel (IMACS),
and arbitrary in the middle panel (DEIMOS).
Two vertical arrows in the middle panel 
indicate the wavelengths of {\sc [Oii]} $\lambda\lambda$3726,3729 doublet 
from a $z=1.5$ {\sc [Oii]} emitter.
\label{fig:blob_2dspec_1dspec}}
\end{figure}

\subsection{Spectroscopic Confirmation}
\label{sec:spectroscopic_confirmation}

We carried out spectroscopic follow-up observations
with Keck/DEIMOS and Magellan/IMACS.
The DEIMOS observations were conducted 
with the 830G grating and the GG495 filter
in the non-photometric night on 2007 November 5.
The IMACS observations were made 
in the nod-and-shuffle mode
with the 150 l mm$^{-1}$ grism and the GG455 filter
under the photometric night on 2007 November 11.
The on-source exposure times of our DEIMOS and IMACS
observations were 10800 and 15600 seconds
in the $0''.8-1''.0$ and $0''.5-0''.8$ seeing conditions,
respectively.
We chose $1''.0$ for the slit width in these observations.
The DEIMOS slit position is plotted in Figure \ref{fig:z7blob_images}.
The slit position of IMACS is the same as that of DEIMOS,
but the position angle is slightly rotated by -1.1 deg. 
The spectral coverages of DEIMOS and IMACS 
are 
$5700-9500$\AA\ and $4500-9700$\AA, respectively.
The spectral resolution of the DEIMOS data at $9200$\AA\
is $R\simeq 3600$, while the one of the IMACS data is $R\simeq 700$.
We have reduced our spectra with spec2d
\footnote{
The analysis pipeline used to reduce the DEIMOS data 
was developed at UC Berkeley with support from NSF grant 
AST-0071048.
}
and COSMOS pipelines
for DEIMOS and IMACS data, respectively.
Both spectra have a strong single line 
with no detectable continuum.
The line-center wavelength of the single line
is $9232.7$\AA\ from our DEIMOS spectrum,
which coincides with the measurement 
from our IMACS spectrum ($9233.5$\AA) within 1\AA.
Figure \ref{fig:blob_2dspec_1dspec} presents our spectra
in the wavelength around this single line.
We have confirmed that the spectra show no signatures of an {\sc [Oiii]} 5007 emission line
(at $\simeq 7044$\AA) and an {\sc [Oii]} 3727 emission line (at $\simeq 5243$\AA) from a $z=0.407$ 
H$\alpha$ emitter or an {\sc [Oii]} emission line (at $\simeq 6872$\AA) from a 
$z=0.844$ {\sc [Oiii]} emitter,
and found that this object is neither a foreground H$\alpha$ nor 
{\sc [Oiii]} emitter. 
We cannot distinguish between an {\sc [Oii]} emitter
at $z=1.477$ and a Ly$\alpha$ emitter at $z=6.595$
from a detection of the other emission line, because 
our spectra do not cover a wavelength that would have another 
strong emission line such as Ly$\alpha$, {\sc [Oii]}, {\sc [Oiii]}, and H$\alpha$.
However, the DEIMOS spectrum has
a FWHM spectral resolution of 2.6\AA\ that would have enabled us 
to identify an {\sc [Oii]} $\lambda\lambda$3726,3729 doublet at $z=1.48$
with a separation of 6.9\AA\
(vertical arrows in the middle panel of Figure \ref{fig:blob_2dspec_1dspec}). 
Our DEIMOS spectrum confirms no such signature of
{\sc [Oii]} doublet, but a clear asymmetric line profile with an extended red wing 
that is typical for a high-$z$ Ly$\alpha$ line. 
We measure the skewness, $S$, and the weighted skewness, $S_w$,
defined by \citet{kashikawa2006}. 
We obtain $S=0.685\pm 0.007$ and $S_w=13.2\pm 0.1$ for our line.
Since the average values of $z\sim 6.5$ LAEs
are $S=0.542\pm 0.007$ and $S_w=11.5\pm 0.2$ \citep{kashikawa2006}, 
the line shape of our object is similar to (or more positively skewed than) 
the average. If this line were an {\sc [Oii]} doublet, 
the line shape would be negatively skewed.
Thus, we conclude that this object is a real LAE at $z=6.595$ with
a clear red wing in the asymmetric Ly$\alpha$ line.
This is the spectroscopic confirmation of the giant LAE, Himiko, at $z=6.595$.
The number density corresponding to this giant LAE is only 
$1.2\times 10^{-6}$ comoving Mpc$^{-3}$ at $z=6.6$
We have identified the rare object near the reionization epoch.

We estimate Ly$\alpha$ flux, $f({\rm Ly}\alpha)$, and
rest-frame equivalent width, $EW_0$, to be
$f({\rm Ly}\alpha)=7.9 \pm 0.5 \times 10^{-17}$ erg s$^{-1}$ cm$^{-2}$ and
$EW_0=100^{+302}_{-43}$ \AA\ with $z'$ and $NB921$-band photometry
in the same manner as \citet{ouchi2008}. The corresponding Ly$\alpha$
luminosity, $L({\rm Ly}\alpha)$, is $L({\rm Ly}\alpha) = 3.9 \pm 0.2 \times 10^{43}$
erg s$^{-1}$.
To check our estimation, we derive $f({\rm Ly}\alpha)$
from the IMACS spectrum that were taken under the photometric condition.
Applying a slit-loss correction,
we obtain $f({\rm Ly}\alpha)=11.2\pm 3.6\times 10^{-17}$ erg s$^{-1}$ cm$^{-2}$
corresponding to $L({\rm Ly}\alpha)=5.6\pm 1.8 \times 10^{43}$ erg s$^{-1}$.
Although the line-flux value from the IMACS spectrum includes the large error,
the line flux from spectroscopy agrees with the one from photometry
within the $1\sigma$ error.
Since the IMACS spectrum shows no continuum above the detection limit,
$EW_0$ cannot be derived from the spectrum.
We summarize spectroscopic as well as various properties 
of this object in Table \ref{tab:properties}.

\begin{deluxetable}{lc}
\tabletypesize{\scriptsize}
\tablecaption{Properties of Himiko
\label{tab:properties}}
\tablewidth{0pt}
\tablehead{
\colhead{Quantity} &
\colhead{Measurement}
}
\startdata
Redshift($z$) & 6.595\\
Skewness of Ly$\alpha$ ($S$) & $0.685\pm 0.007$\\
Weighted Skewness of Ly$\alpha$ ($S_w$) & $13.2\pm 0.1$\\
Isophotal Area\tablenotemark{1}($NB921$) & 5.22 arcsec$^2$ \\
Isophotal Area\tablenotemark{1}($z'$)    & 1.88 arcsec$^2$ \\
Major Axis\tablenotemark{2}($NB921$) & $3.1''$ \\
Major Axis\tablenotemark{2}($z'$) & $2.0''$ \\
Ly$\alpha$ Surface Brightness & $1.51 \times 10^{-17}$erg s$^{-1}$cm$^{-2}$arcsec$^{-2}$\\ 
$NB921$ Surface Brightness & 25.5 mag arcsec$^{-2}$\\
$f({\rm Ly}\alpha)$\tablenotemark{3} & $7.9 \pm 0.5 \times 10^{-17}$ erg s$^{-1}$ cm$^{-2}$\\
$L({\rm Ly}\alpha)$\tablenotemark{3} & $3.9 \pm 0.2 \times 10^{43}$ erg s$^{-1}$ \\
$M(1250)$\tablenotemark{4} & $-21.67 \pm 0.73$ mag \\
$M_B$\tablenotemark{4} & $-22.83 \pm 0.27$ mag \\
Ly$\alpha$ Line Width (FWHM) & $251\pm 21$ km s$^{-1}$\\
Rest-Frame EW ($EW_0$) & $100^{+302}_{-43}$ \AA\ \\
Stellar Mass & $3.5^{+1.5}_{-2.6}\times 10^{10}M_\odot$\\
SFR from SED fit\tablenotemark{5} & $>34 M_\odot$yr$^{-1}$ \\
SFR from UV\tablenotemark{5} & $25^{+24}_{-12} M_\odot$ yr$^{-1}$  \\
SFR from Ly$\alpha$\tablenotemark{5} & $36\pm 2 M_\odot$ yr$^{-1}$ \\
Specific SFR & $>1.6\times 10^{-9}$yr$^{-1}$\\
Number Density\tablenotemark{6} & $1.2\times 10^{-6}$ Mpc$^{-3}$\\
\enddata
\tablenotetext{1}{The isophotal areas are defined as pixels with values above 
the $2 \sigma$ sky fluctuation; $26.8$ and $27.3$ mag arcsec$^{-2}$ in the $NB921$ and $z'$ images, 
respectively.}
\tablenotetext{2}{The maximum size of the $2\sigma$ isophotal area. 
For isophotal areas above the $3\sigma$ sky fluctuation, we obtain $2''.7$ and $1''.2$ 
in the $NB921$ and $z'$ images, respectively.}

\tablenotetext{3}{The Ly$\alpha$ flux and luminosity from the photometric measurements.}
\tablenotetext{4}{The rest-frame 1250\AA, $M(1250)$, and $B$-band, $M_B$, magnitudes. 
Since the $3.6\mu$m band observes
the rest-frame 4180-5168\AA\ which is very close to the bandpass of $B$ band,
no k-correction is applied to the $3.6\mu$m-band magnitude.}
\tablenotetext{5}{The SFRs estimated from the SED fitting,
the UV continuum, and the Ly$\alpha$ luminosity. The SFRs of 
UV and Ly$\alpha$ are not corrected for dust extinction. 
See \S \ref{sec:stellar_population} for more details.
}
\tablenotetext{6}{The comoving number density corresponding to Himiko.
}
\end{deluxetable}

\section{A Giant LAE at $\lowercase{z}=6.595$}
\label{sec:giant_LAE}

\subsection{AGN Activity} 
\label{sec:agn}

We investigate the AGN activity in our object.
Our spectra show no {\sc Nv} 1240 line at 9418\AA\ as well as
no broadening of Ly$\alpha$. Moreover, there is no counterpart
in our Spitzer MIPS image as well as the XMM-Newton X-ray and 
VLA radio catalogs \citep{ueda2008,simpson2006}
whose detection limits are $m(24\mu{\rm m})=19.8$,
$f(0.5-2{\rm keV})=6\times 10^{-16}$ erg cm$^{-2}$ s$^{-1}$, and
$f(1.4{\rm GHz})=100\mu$Jy, respectively.
\begin{figure}
\epsscale{1.2}
\plotone{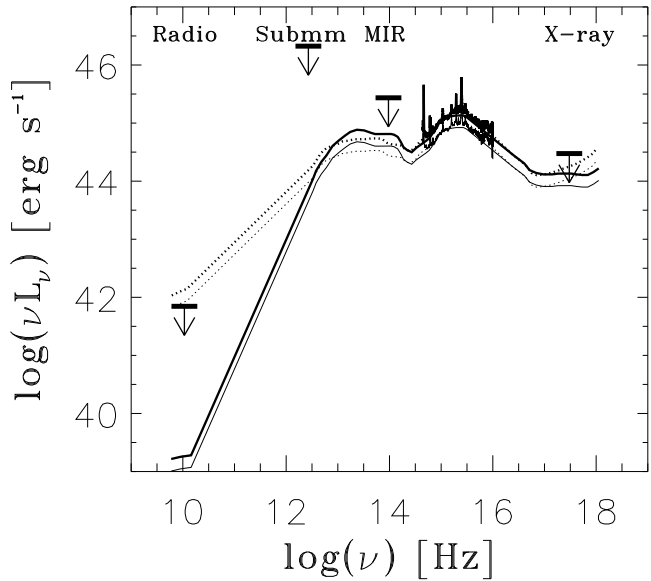}
\caption{
The upper limits to the spectral energy distribution
(SED) of our object in the rest frame.
The thick bars with an arrow
indicate the upper limits
that correspond to
the VLA (Radio), SCUBA (Submm),
Spitzer/MIPS (MIR), and
XMM-Newton (X-ray) results
from left to right.
Solid and dotted curves are the templates of radio-quiet
and radio-loud AGNs taken from
\citet{elvis1994}, \citet{telfer2002}, and \citet{richards2003}.
The thick and thin curves represent the template SEDs
normalized with Ly$\alpha$ and rest-frame optical
luminosities of our object, respectively.
\label{fig:z7blob_multibandsed}}
\end{figure}
Our object is not also detected 
in the SCUBA Half-Degree Extragalactic Survey 
(SHADES; \citealt{mortier2005}) data.
We find no significant signal at 
the location of our object
in the submm $850\mu$m map, and obtain the 
$3\sigma$ upper limit flux of $S(850\mu m)<12$ mJy.
We present these upper limits in Figure \ref{fig:z7blob_multibandsed},
together with the AGN templates used in \citet{ouchi2008}.
The AGN templates are
the average radio quiet/loud quasars \citep{elvis1994},
and a UV-optical spectrum of \citet{telfer2002} and \citet{richards2003}
that is normalized to the average spectral energy distribution (SED) of 
\citet{elvis1994}.
We choose the amplitudes ($\log \nu L_\nu$) of these AGN templates 
that match to Ly$\alpha$ (thick lines) and optical luminosities 
(thin lines) of our object. Here, the Ly$\alpha$ luminosity of 
the templates is defined by the Ly$\alpha$ flux that enters
into the bandpass of our narrow-band ($\lambda_{\rm rest}=1207-1224$\AA).
The optical luminosity is the one at rest frame $\simeq 4700$\AA\ 
that corresponds to the Spitzer $3.6\mu$m band.
Figure \ref{fig:z7blob_multibandsed} indicates that 
our object is probably not a typical radio-loud AGN,
although the upper limit of radio band is marginally
below the templates. On the other hand,
none of our multi-wavelength data can 
place constraints on the existence of 
radio-quiet AGN. Thus, we cannot discard the
possibility that our object is an AGN.
It is also true that there are no 
positive signatures of AGN activities
in our spectroscopic and multi-wavelength data.

Although it is not clear whether this object has
an AGN, we investigate how much our broad-band
photometry is impacted by prominent nuclear emission lines
in the case of AGN.
Strong {\sc [Oiii]} and H$\alpha$ of AGN
would enter into the $3.6\mu$m and $4.5\mu$m bands, respectively,
which may boost the magnitudes in these broad bands.
We estimate the possible contributions from these strong lines, 
assuming the flux ratios of
$f_{{\rm Ly}\alpha}/f_{\rm [OIII]}=4$ for a type II AGN
(\citealt{mccarthy1993}; due to no Ly$\alpha$ broadening) 
and $f_{{\rm Ly}\alpha}/f_{\rm {\rm H}\alpha}=8.7$
for the case B recombination \citep{brocklehurst1971}. 
We, thus, obtain 26.3 and 27.0 magnitudes in $3.6\mu$m and $4.5\mu$m bands, respectively.
Since the magnitude of our object is $24.0$ in the $3.6\mu$m band, the flux contributed 
from strong lines of AGN would be about one-order of magnitude smaller than 
the brightness of our object. Even with an AGN,
the magnitudes of $3.6\mu$m and $4.5\mu$m bands 
would include a negligible contribution from strong emission lines.
It should be noted that these flux contributions from strong lines could
be underestimated, in the case where our Ly$\alpha$ flux is very strongly absorbed.

\subsection{Possibility of Gravitational Lensing} 
\label{sec:gravitational_lensing}
Strong lensing is more common among very high-redshift objects
because of increasing lensing optical depth with increasing
redshift (e.g., \citealt{hilbert2007}; \citealt{hilbert2008}) and larger effect of
magnification bias for very high-redshift objects (e.g., \citealt{wyithe2002}). 
Moreover, the extended nature of Himiko may make
the lensing interpretation of this object plausible.
In this section, we investigate the possibility of gravitational
lensing.
First, we use the catalog of \citet{vanbreukelen2006}
which shows cluster candidates at $z=0.5-1.5$ in this field. 
The estimated masses of these clusters range between
$5\times 10^{13}-3\times 10^{14} M_\odot$. Our object
is separated from the center of the closest cluster candidate 
by $\simeq 5'.8$.
Due to this large separation, the magnification by these clusters
is negligible. 

Next, we investigate the possibility of galaxy-galaxy lensing.
Figure \ref{fig:z7blob_images} indicates that
this source may have two peaks in the $z'$ image,
and a smooth profile in the $NB921$ image.
From a visual inspection, the $3.6\mu$m-band profile 
would appear to be slightly elongated.
However, our object is not bright enough in the $3.6\mu$m band 
to distinguish between profiles of a real extended source and
a point source with outskirts made by peaks of background fluctuations.
We carry out profile fitting to our $z'$ and $NB921$ images
with GALFIT \citep{peng2002}. We fit two profiles whose flux amplitudes
and positions are free parameters.
We find that fitting with two PSF profiles leaves 
large residuals, and that the profile of our object
is well fit by 
two circular exponential disks with a half-light radius of $R_{\rm hl}=0''.3$
which are separated by $1''.1$.
The positions of two components are 
determined in the $z'$-band image that shows 
the possible two peaks,
and presented with green arrows in Figure \ref{fig:z7blob_images}.
On the other hand, the two-exponential disk models
can reproduce not only the $z'$-band, but also 
the $NB921$-band profiles.
The positions of the best-fit models in the $NB921$ image
are not the same as those in the $z'$ image, 
but the differences of the positions
are only $0''.2-0''.3$.
Because the photometric uncertainties in the $z'$ band are large,
it is not clear whether two peaks really exist or whether
the positions of two peaks are different between the $z'$ and $NB921$
images.
Nevertheless, we refer to the positions of east and west components 
determined in the $z'$ image as $position$ 1 and 2, respectively.
The brightness ratio, $\Delta m$, 
of $position$ 1 ($m_1$) and 2 ($m_2$) components 
are $\Delta m \equiv m_2-m_1=0.38 \pm 0.38$ and $0.65 \pm  0.10$
in $z'$ and $NB921$ bands, respectively.
There are no significant differences between $\Delta m$ of $z'$ and $NB921$ bands. 
We cannot reject the possibility that these two components have
the same color, and that these components are an identical lensed object.
Thus, our profile fitting does not constrain
the possibility of lensing due to the large photometric uncertainties of $\Delta m$.

\begin{figure}
\epsscale{1.1}
\plotone{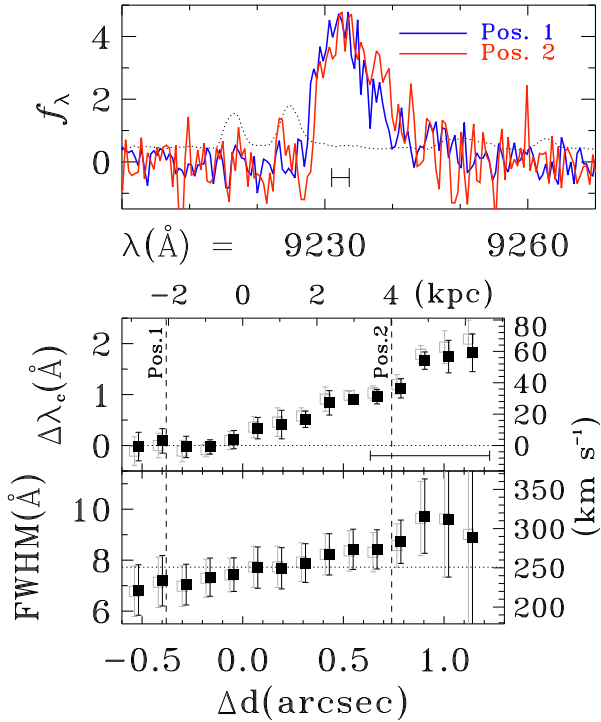}
\caption{
$Top$: DEIMOS spectra at $position$ 1 (blue) and
2 (red). The dotted line represents the background sky. The
instrumental FWHM is indicated with the bar below the emission
line. 
$Bottom$: The line-center offset (upper panel) and 
FWHM line width (lower panel) as a function of position. 
The $\Delta \lambda_c$ is defined
by $\Delta\lambda_c=\lambda_c - \lambda_c(\Delta d=0)$,
where $\lambda_c$ is the line-center wavelength.
The definition of $\Delta d$ is the same as 
that of the top panel in Figure \ref{fig:blob_2dspec_1dspec},
and $\Delta d=0$ corresponds to the Ly$\alpha$ source center.
The gray open squares are the direct measurements,
while the black squares are the best estimates after the
correction (see the text). The size of spatial binning ($0".6$)
is shown with the bar on the right side.
The dotted lines present the values at $\Delta d=0$. 
The $positions$ of 1 and 2 are indicated with the dashed lines.
The right-hand vertical axis ticks the corresponding velocity.
The FWHM-line width is corrected for the  
instrumental broadening with the assumption of a Gaussian profile.
\label{fig:z7blob_specdiff}}
\end{figure}

Figure \ref{fig:z7blob_specdiff} presents DEIMOS spectra 
at $position$ 1 and 2 whose extraction-aperture size
is $0''.6$ along the slit.
The skewness and the weighted skewness of the emission line 
are estimated to be
$(S, S_w)=(0.846\pm 0.018, 17.3\pm 0.4)$ and $(0.502\pm0.023, 9.3\pm 0.4)$
in the spectra
at ($position$ 1, $position$ 2).
These two spectra show an asymmetric line with
$S$ and $S_w$ which are comparable with 
the average values of $z\sim 6.5$ LAE
(see \S \ref{sec:spectroscopic_confirmation}).
Thus, both of these components reside at high-$z$. 
The bottom panel of Figure \ref{fig:z7blob_specdiff} presents
line-center offset and line width as a function of distance 
along the DEIMOS slit. 
We measure the line center and width by Gaussian-profile fitting.
Note that the $NB921$ (or Ly$\alpha$) surface-brightness distribution
is not homogeneous within the slit.
In fact, Figure \ref{fig:z7blob_images} implies that,
from $position$ 1 to 2,
the profile center would shift towards the direction of red spectrum (magenta
arrows in Figure \ref{fig:z7blob_images}).
We estimate the biases raised by this profile inhomogeneity
with the $NB921$ image which has a seeing size comparable to our DEIMOS spectrum.
We measure changes of the profile's center and
standard deviation within the slit as a function of
slit position in the $NB921$ image. We calculate
the correction factors in wavelength based on 
these spatial changes of $NB921$ profile, and
apply these correction factors to the original 
measurements of line-center offset and line width.
The bottom panel of Figure \ref{fig:z7blob_specdiff} shows 
the corrected values (filled squares), together with
the original measurements (open squares).
We confirm that those biases are not large enough 
to alter general trends in line-center offset and line width. 
Figure \ref{fig:z7blob_specdiff} indicates that
the Ly$\alpha$-line center shifts 
gradually by $\simeq 30$ km s$^{-1}$ between 
these two components. Thus, these components are not 
sources of an identical object produced by lensing because of this line shift.
It indicates that it is unlikely that they are gravitationally 
lensed objects. Moreover, following the method introduced 
in \S 2.4.2 of \citet{pindor2006},
we estimate the minimum brightness of lensing galaxy at $z<4$
to be $K=23.2$ based on our source redshift of $6.6$ and 
image separation of $1''.1$.
Figure \ref{fig:z7blob_images} presents no nearby bright sources
with $K\lesssim 23.2$ to which can be ascribed a lensing object.
There remains a special case where a very red foreground object 
with a $3.6\mu$m-band detection and no optical-NIR counterparts 
magnifies an inhomogeneous LAE 
with spatially different magnification factors.
However, there are little chance coincidences
in the precise alignment of such an extremely-red lensing object.
We thus conclude that our LAE is not likely to be a lensed source, 
but an intrinsically extended object. The size of the extended Ly$\alpha$ nebula
is $\gtrsim 17$ proper kpc at $z=6.595$ which is estimated from the major axis of 
the isophotal area in the $NB921$ image ($\gtrsim 3''.1$; \S \ref{sec:photometric_identification}).

\subsection{Kinematics of Resonance Ly$\alpha$ Line} 
\label{sec:kinematics}

Because Ly$\alpha$ is a resonance line strongly 
scattered or absorbed by gas and dust,
it is difficult to determine the gas kinematics 
by Ly$\alpha$ observations alone.
However, it is known that Ly$\alpha$ line is a useful probe of 
gas inflow/outflow and cosmic reionization with
detailed modeling \citep{tapken2007,dijkstra2007a}.
First, we obtain the line width of $v_{\rm FWHM}=251\pm 21$ km s$^{-1}$ from the DEIMOS spectrum 
in the extraction aperture of $1''.2\times 1''.0$ (i.e. $6.5\times 5.4$ proper kpc$^2$)
around the Ly$\alpha$ source center which corresponds to
$\Delta d=0$ in Figure \ref{fig:z7blob_specdiff}. The line width
is corrected for the instrumental broadening
with the assumption of a Gaussian profile.
The line width can be twice as large as this value,
if a blue half of Ly$\alpha$ is completely 
absorbed by IGM with no effects of the Ly$\alpha$ damping wing.
From further inspection of the line-center offset and line width
along our DEIMOS slit (Figure \ref{fig:z7blob_specdiff}), 
we find that the line-center velocity of 
Ly$\alpha$ increases by 
$\Delta v \simeq 60$ km s$^{-1}$  
from east to west in a range of $\sim 2''$ ($D = 10$ proper kpc).
Note that this small velocity offset is larger than
the sizes of their error bars.
\footnote{
Although the instrumental spectral resolution is $R\sim 3600$
corresponding to $v_{\rm FWHM}\sim 80$ km s$^{-1}$, the uncertainties of
line centering by Gaussian fitting is as small as $\sim 5-10$ km s$^{-1}$ (See error bars in
Figure \ref{fig:z7blob_specdiff}).
}
On the other hand, there are no significant changes of line width beyond the sizes of 
their error bars, 
although our spectrum implies an increase by $\sim 50$ km s$^{-1}$ from east to west.

\subsection{Stellar Population and Mass} 
\label{sec:stellar_population}

\begin{figure}
\epsscale{1.2}
\plotone{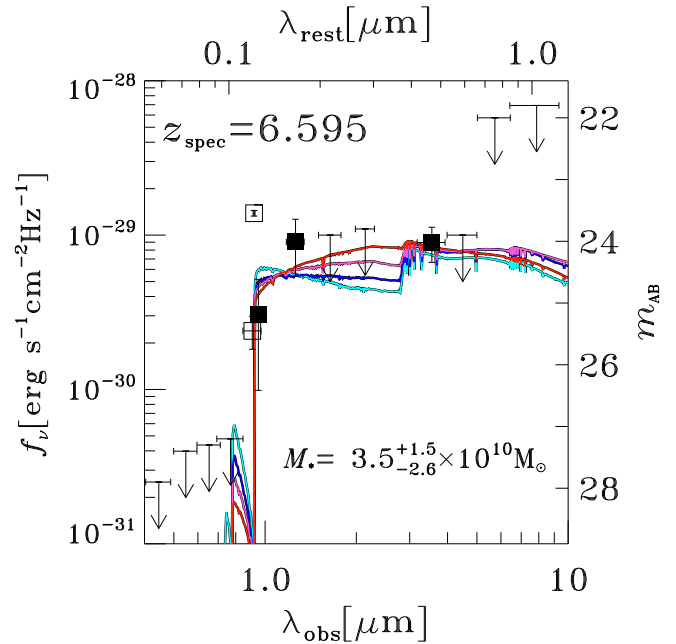}
\caption{
Spectral energy distribution (SED) of the giant LAE, Himiko. 
The squares represent the total fluxes of this object
taken from Table \ref{tab:photometry}.
The open symbols are data points that are not used
for the SED fitting, since these bands are contaminated
by the strong Ly$\alpha$ emission line.
The cyan, blue, magenta, and red lines show 
the best-fit stellar synthesis models
with $E(B-V)=0.0$, 0.3, 0.6, 0.9, respectively,
under the assumptions of constant star formation
with a fixed metallicity of $Z=0.02Z_\odot$.
\label{fig:z7blob_sed}}
\end{figure}

We carry out $\chi^2$ fitting of stellar synthesis models 
to the SED of this object
based on total fluxes at the observed-frame of $0.9-8.0\mu$ m
(Table \ref{tab:photometry}).
We use our best estimates of the total fluxes with the associated $1\sigma$ error 
for all of photometry points including those below the detection limits. 
Since our $z'$-band photometry is contaminated by Ly$\alpha$ emission
and Gunn-Peterson trough, we estimate the emission-free continuum magnitude
at $9500$\AA, $m_{0.95}$.
We obtain $m_{0.95}=25.18\pm 0.73$ based on $NB921$- and 
$z'$-band photometry by the method similar to that of \citet{shimasaku2006}, which 
takes account the contributions of the Ly$\alpha$ line and 
IGM absorption \citep{madau1995} with the response curves of 
$NB921$ and $z'$ filters.
We use the stellar synthesis models of \citet{bruzual2003} with dust attenuation
of \citet{calzetti2000}. 
Applying models of constant and exponentially-decaying star-formation
histories with
sets of metallicity in $Z=0.02-1.0Z_\odot$,
we search for the best-fit model in a parameter space of
$E(B-V)=0-1$ and age$=1-810$ Myr, where the upper limit of 
stellar age is the cosmic age at $z=6.595$.
First, we assume the constant star-formation with a fixed metallicity of 
$Z=0.02Z_\odot$.
Figure \ref{fig:z7blob_sed} presents 
the SED and the best-fit models.
We find that the best-fit model has a
stellar mass of $M_*=3.5^{+1.5}_{-2.6}\times 10^{10}M_\odot$
(i.e. $0.9-5.0\times 10^{10}M_\odot$)
with a reduced $\chi^2$ of 0.84.
Because very weak photometric constraints are given in the rest-frame near UV
($\sim 0.2-0.3\mu$m) critical to resolving the degeneracy
between extinction and age,
we obtain no meaningful measurements on extinction 
and stellar age within the $\simeq 1\sigma$ error.
For examples, the sets of allowed parameters
are (E[B-V], age[Myr])= 
$(0.0, 810)$,
$(0.3, 200)$,
$(0.6, 29)$, and
$(0.9, 3)$.
On the other hand, our estimate of stellar mass 
($0.9-5.0\times 10^{10}M_\odot$)
has a moderate reliability
due to the determinations of precise spectroscopic redshift
and the rest-frame optical photometry 
on which stellar mass primarily depends.
Because the sets of allowed extinction and age parameters
cancel out the variance of mass-to-luminosity ratio ($M/L$),
stellar mass is obtained with a much smaller error than
extinction and age (Papovich et al. 2001).
It should be noted that this object is the most distant spectroscopically-confirmed
galaxy whose stellar mass is constrained.
Next, we change the metallicity to $Z=0.2-1.0Z_\odot$
and star-formation history to an exponential-decay time scale of
$\tau=1-100$ Myr. The stellar mass is estimated to be 
$1.7-3.2\times 10^{10} M_\odot$ and $0.4-4.8\times 10^{10} M_\odot$ 
for the best-fit values and the $1\sigma$ error ranges, respectively.
We find that the general behaviors of the fitting is the same, 
and that the best-fit values and the $1\sigma$ error ranges of stellar mass
agree with those from the first assumptions ($0.9-5.0\times 10^{10}M_\odot$) 
within a factor of $\simeq 2$.

\citet{mobasher2005} report a very massive galaxy at $z_{\rm phot}=6.5$
with a stellar mass of $6 \times 10^{11} M_\odot$ based on a photometrically-selected galaxy.
However, it has been pointed out that \citeauthor{mobasher2005}'s object may be a lower redshift 
($z\sim 2$) dust-enshrouded starburst galaxy, based on a reanalysis of its photo-$z$ \citep{dunlop2007} 
and the detection of polycyclic aromatic hydrocarbon emission features \citep{chary2007}.
On the other hand, \citet{labbe2006} study $z$-dropout galaxies with a photo-$z$ of $z\sim 7$
in the Hubble Ultra Deep Field (HUDF), and find that these galaxies have stellar masses of 
$0.1-1 \times 10^{10}M_\odot$. This mass range touches the lowest-mass limit of 
our object ($0.9-5.0\times 10^{10}M_\odot$).
If the \citeauthor{labbe2006}'s photo-$z$ objects are real high-$z$ galaxies,
our object is likely to be a more stellar-massive galaxy than those found 
in the small area of HUDF. 
Similarly, \citet{egami2005}, \citet{chary2005}, and \citet{schaerer2005} have estimated stellar masses
of gravitationally-lensed galaxies at $z\sim 7$ behind Abell clusters 
to be $\sim 10^{8}-10^{9} M_\odot$, which is $\sim 1-2$ orders of magnitude smaller than
the stellar mass of our object.

Because $E(B-V)$ cannot be constrained,
we can only obtain the lower-limit of star-formation rate
of $SFR>34 M_\odot$yr$^{-1}$ from the SED fitting,
which is given in the case of $E(B-V)=0$.
This lower limit from the SED fitting is consistent with
the SFRs estimated from the UV continuum of $m_{0.95}$ (SFR=$25^{+24}_{-12} M_\odot$ yr$^{-1}$)
and from the Ly$\alpha$ luminosity (SFR=$36\pm 2 M_\odot$ yr$^{-1}$)
with no dust-extinction corrections via formulae of
\citet{madau1998} and \citet{kennicutt1998} + case B recombination,
respectively.

The SSFR of our object is 
$SSFR>1.6\times 10^{-9}$yr$^{-1}$
at the stellar mass of $0.9-5.0 \times 10^{10}M_\odot$.
In the plane of SSFR vs. stellar mass, our lower limit of SSFR 
is comparable to LBGs at $z\simeq 2-3$
(see, e.g., \citealt{castroceron2008}). 
However, the stellar mass of our object is, at least, 
one-order of magnitude larger than
that of the averaged (stacked) LAEs at $z\sim 3$ 
\citep{nilsson2007,gawiser2007,lai2008}.
Our derived stellar mass is more similar to those of luminous LAEs at $z=3.1-5.7$
that are bright enough to be identified individually in infrared images
(\citealt{lai2007,finkelstein2008}; Ono et al. in prep; cf. very faint LAEs
in the HUDF by \citealt{pirzkal2007}).

We calculate four statistical measurements from
our object's SFR, stellar mass, and
number density listed in Table \ref{tab:properties},
and compare with those obtained by the other studies. 
The comparisons are useful to check 
how our object plays a role in the average volume of 
the Universe at $z\sim 7$.
(i) We estimate the lower limit of UV luminosity function (LF) to be 
$\gtrsim 1.2\times 10^{-6}$ mag$^{-1}$ Mpc$^{-3}$ 
at $M_{\rm UV}\simeq -21.3$. 
This limit is consistent with $z\sim 7$ UV LF of 
\citet{bouwens2008}.
(ii) The lower limit of cosmic SFR is 
$>4.3\times 10^{-5} M_\odot$ yr$^{-1}$ Mpc$^{-3}$,
which is one to two orders of magnitude smaller 
than that obtained by \citet{bouwens2008}.
Thus, this limit is consistent with the estimate of 
\citet{bouwens2008}. 
The small contribution to the cosmic SFR leaves 
the possibility that significantly luminous objects like 
the one we discuss here could not be the major contributors of 
cosmic reionization at $z\sim 7$
similar to at $z\sim 6$ \citep{yan2004}.
(iii) The lower limit of stellar-mass function is 
$\log \rho_* > -5.9$ Mpc$^{-3}$ dex$^{-1}$ at 
$\log(M_*)\simeq 10.5$. This limit is much lower 
than stellar mass function of 
$z\sim 5$ dropouts given by \citet{mclure2008}. 
It indicates that our lower limit
is compatible with the scenario of hierarchical structure formation,
because a similarly massive system is more abundant
at the recent epoch of $z\sim 5$ than at $z\sim 7$.
(iv) The lower limit of stellar-mass density is 
$>4.4\times 10^4 M_\odot$ Mpc$^{-3}$,
which is consistent with that obtained in the HUDF
($1.6\times 10^6 M_\odot$ Mpc$^{-3}$; \citealt{labbe2006}).
All of these four statistical measurements corresponding
to our object fit in the average properties of 
the Universe at $z\sim 7$.

\section{Discussion}
\label{sec:discussion}

\subsection{Comparisons with Ly$\alpha$ Blobs at $\lowercase{z}\sim 3$} 
\label{sec:comparisons}

\begin{figure}
\epsscale{1.2}
\plotone{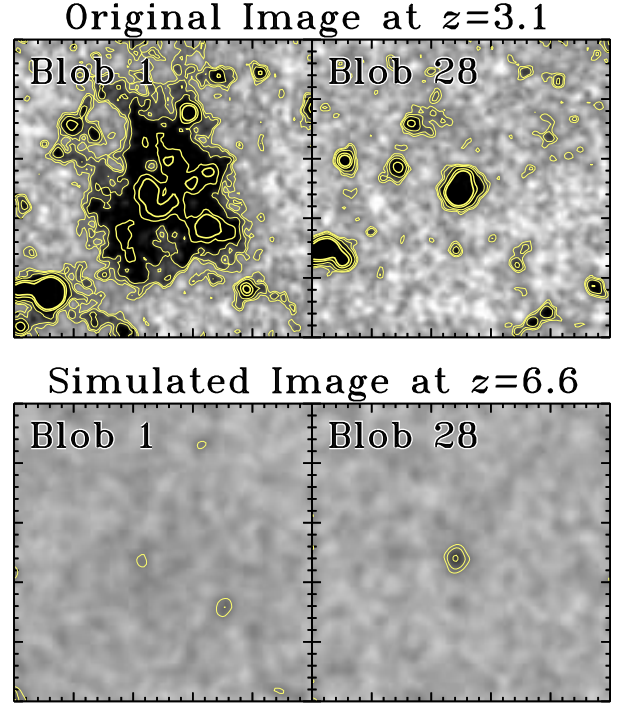}
\caption{
Original narrow-band images of blob 1 and 28 at $z=3.1$
from \citet{matsuda2004} (top panels) and simulated
narrow-band images at $z=6.6$ (bottom panels).
The size of each panel is $25''\times25''$.
The intensity contours are presented with yellow lines.
The contours represent (2, 3, 5, 10, 15) $\sigma$ 
levels of sky fluctuations,
where the $2\sigma$ sky fluctuations correspond to
$28.0$ and $26.8$ mag arcsec$^{-2}$
in the original images \citep{matsuda2004} and the
simulated images, respectively.
Note that the absolute values of contour levels
in the simulated images
are the same as the yellow contours in the $NB921$ image 
of Figure \ref{fig:z7blob_images}.
In the original images, blob 1 is the large diffuse extended
source, while blob 28 is the compact source at the center.
Note that the original and simulated images include
foreground and background sources neighboring the blobs in the
$25''\times25''$ areas.
\label{fig:z7blob_simu}}
\end{figure}

We compare properties of our giant LAE with
those of Ly$\alpha$ blobs found at $z\simeq 3$.
Because objects experience more severe surface-brightness dimming at $z=6.6$
than at $z\simeq 3$, careful comparisons are needed.
We produce simulated $NB921$ images of blob 1 of \citet{steidel2000}
and blob 28 of \citet{matsuda2004} redshifted to $z=6.595$ 
based on the narrow-band images of \citet{matsuda2004}.
We choose the blob 1 and 28, because they have 
the brightest Ly$\alpha$ luminosity and the highest surface brightness, respectively,
in \citeauthor{matsuda2004}'s catalog.
We carry out the simulations in the same manner as \citet{saito2006}, 
but with an improved random noise whose amplitude exactly matches
to those of real $NB921$ image.
Figure \ref{fig:z7blob_simu} presents the simulated $NB921$ images,
together with the original narrow-band images of \citet{matsuda2004}.
The simulated image of blob 1 indicates
that our observations would miss $\simeq 95$\% of 
blob 1's $L({\rm Ly}\alpha)$, and that 
no diffuse Ly$\alpha$ nebula
could be identified. The simulated blob 1 has
($NB921$[total], $A_{\rm iso}$, $\left <SB\right >$)
$=$ ( $25.5$ mag, $0.8$ arcsec$^2$, $26.7$ mag arcsec$^{-2}$).
We plot this simulated object in Figure \ref{fig:mag_isoarea_NB921LAE}.
The simulated blob 1 is indistinguishable from the cloud of normal LAEs
in the planes of $A_{\rm iso}$ vs. $NB921$ and $A_{\rm iso}$ vs. $\left< SB \right>$.
On the other hand, the simulated blob 28 shows
($NB921$[total], $A_{\rm iso}$, $\left <SB\right >$)
$=$ ( $24.1$ mag, $3.7$ arcsec$^2$, $26.3$ mag arcsec$^{-2}$).
This simulated object is recognizable in Figure \ref{fig:mag_isoarea_NB921LAE}.
However, this simulated object is more similar to  
the other LAEs with the second and third largest $A_{\rm iso}$
than to our object. There exist remarkable differences in $A_{\rm iso}$
and $\left <SB\right >$ (or $NB921$ magnitude) between our object and the simulated objects.
In other words, the large $A_{\rm iso}$ and the high $\left <SB\right >$
of our object at $z=6.6$ cannot be realized 
even for blob 28 whose very high surface brightness would
minimize the effect of cosmological surface brightness dimming.
It implies that our object might be a population that
has not yet been identified at $z\simeq 3$.

Although the Ly$\alpha$ brightness and profile
of our object seem different from those of Ly$\alpha$ blobs 
at $z\simeq 3$, the other properties of our object 
show both similarities and differences.
The line width of our object is comparable 
to some of the extended LAEs
presented in \citet{matsuda2006} 
in the plane of FWHM(Ly$\alpha$) vs. isophotal area,
although the threshold of isophotal area is different from ours.
On the other hand, there are no objects with such 
a luminous Ly$\alpha$ line in the range of $v_{\rm FWHM}=100-900$ km s$^{-1}$
in the catalog of \citet{saito2008}.
The velocity width of our object is $251$ km s$^{-1}$ which is
significantly smaller than that of \citeauthor{steidel2000}'s blob 1 
with a broad-velocity width, $\simeq 1000-1500$ km s$^{-1}$ (\citealt{ohyama2003};
see also \citealt{bower2004}). The velocity width of our object
is closer to the one of blob 28 ($v_{\rm FWHM}=362$ km s$^{-1}$; Y. Matsuda in preparation).
The stellar masses of the sub-components shaping \citeauthor{steidel2000}'s blobs 1 and 2 
range from $10^{10}$ to $10^{11}M_\odot$ \citep{uchimoto2008},
which is comparable to ours. Similarly, \citet{smith2008} report
that the stellar mass of their Ly$\alpha$ blob at $z=2.83$
is as massive as $3-4\times 10^{11} M_\odot$. These stellar masses
are comparable to ours within an order of magnitude.
The estimated number density of our object is 
only $1.2\times 10^{-6}$ comoving Mpc$^{-3}$, which
is lower than those of $z\sim 3$ Ly$\alpha$ blobs
found by \citet{matsuda2004} ($3\times 10^{-4}$ Mpc$^{-3}$),
\citet{saito2006} ($1\times 10^{-5}$ Mpc$^{-3}$), and
Yang et al. in preparation ($3\times 10^{-6}$ Mpc$^{-3}$).
This number density is also lower than the upper limit of 
number density of \citeauthor{steidel2000}'s two Ly$\alpha$ blobs
($2\times 10^{-5}$ Mpc$^{-3}$), which are estimated from
the number of objects (2) and 
\citeauthor{matsuda2004}'s survey volume ($1.3\times 10^{5}$ Mpc$^{3}$).
Although these number densities depend on the criteria of 
sample selections and observing fields,
our object at $z=6.6$ would be as rare as (or even rarer than) 
these Ly$\alpha$ blobs at $z\sim 3$.
Regarding the environment, our giant LAE resides in a high density region of 
LAEs (\S \ref{sec:photometric_identification}).
\citet{matsuda2004} find that, at $z=3.1$, the distribution
of their Ly$\alpha$ blobs trace the dense regions of LAEs. 
The similar spatial correlation between our extended LAE and compact LAEs
is also seen at $z=6.6$.

We also compare our object with Ly$\alpha$ nebulae around
high-$z$ radio galaxies (HzRGs; \citealt{miley2008}). 
The Ly$\alpha$ luminosity of our object is comparable to
those of HzRGs ($\sim 10^{43}-10^{44}$ erg s$^{-1}$;
\citealt{mccarthy1991}). On the other hand, our object has a much smaller
FWHM line width of Ly$\alpha$, $251\pm21$ km s$^{-1}$, 
than HzRGs ($\gtrsim 1000$ km s$^{-1}$; \citealt{mccarthy1996}).
However, recent studies have found an HzRG with a strikingly relaxed dynamics
whose Ly$\alpha$ FWHM line width is $<300$ km s$^{-1}$ \citep{villar-martin2007}.
It would be difficult to distinguish between our object
and such a dynamically-relaxed HzRG powering a Ly$\alpha$ halo based on dynamics,
although our object has a radio flux fainter than that of a typical radio-loud AGN 
(Section \ref{sec:agn}).

\subsection{Inferred Kinematics}
\label{sec:inferred_kinematics}

If this object forms a single virialized system 
whose possible two $z'$-band components would be just bright {\sc Hii} regions
in a disk (see \S \ref{sec:nature}),
the dynamical mass from the rotation is naively estimated
to be $M_{\rm rot} \sin(i) \simeq 1\times 10^9 M_\odot$ by 
$M_{\rm rot} \sin(i)= ([v_c \sin(i)]^2 r )/G$ 
with $v_c \simeq \Delta v/2 = 30$ km s$^{-1}$ 
and $r= D/2 = 5$ kpc (\S \ref{sec:kinematics}),
where $i$ and $G$ are
the inclination of a rotating disk and the gravitational constant.
On the other hand, from the 1-dimensional 
velocity dispersion of $\sigma_{v}=v_{\rm FWHM}/2.35=107$ km s$^{-1}$
in a half size of the extraction aperture, $R\sim 6/2=3$ kpc (\S \ref{sec:kinematics}), 
we obtain the mass of random motion
of $M_{\rm rand}=4\times 10^{10} M_\odot$ via 
$M_{\rm rand} = (5/3) (3 \sigma_{v}^{2}) R/G$, 
assuming a uniform sphere.
If the blue half of Ly$\alpha$ is absorbed by
the external IGM, the velocity dispersion and mass are 
$\simeq 214$ km s$^{-1}$ and $M_{\rm rand}=2\times 10^{11} M_\odot$, respectively.
Thus, $M_{\rm rand}$ is 1-2 order(s) of magnitude larger than
$M_{\rm rot}$ in a reasonable range of inclination ($i=10-90^\circ$).
If we consider the smaller radius for the $M_{\rm rand}$ estimate ($R= 3$ kpc)
than that for the $M_{\rm rot}$ estimate ($r= 5$ kpc),
the difference of these dynamical masses defined in a common radius
becomes even larger.
If Ly$\alpha$ reflects dynamics,
our giant LAE would be a system more dominated by random motion
than rotation. Note that these estimates of dynamical masses
depend on the size of spectrum extraction aperture
with the uncertainties of seeing smearing,
and that these results are only true under the
assumptions of the single-virialized system and no significant effects of 
resonant scattering.

If this giant LAE is an outflow object
whose Ly$\alpha$ emission is produced by shock heating (cf. \citealt{dijkstra2008}),
the dynamical time scale required to form the extended Ly$\alpha$ nebula 
is $\simeq 7\times 10^7$ yr, where we assume
the size of the major axis ($\simeq 17$ kpc)
and the typical velocity width ($v_{\rm FWHM}=251$ km s$^{-1}$).
Since this time scale is as long as the cosmic time 
between $z=6.595$ and $7.1$, it would start making an 
ionized-bubble since $z=7.1$ in this shock heating scenario.

\subsection{Nature of the giant LAE}
\label{sec:nature}

The nature of the Ly$\alpha$ nebula of our object is not yet clearly
understood within the currently available observational data.
There are five possible explanations:
(1) halo gas photoionized by a hidden AGN,
(2) clouds of {\sc Hii} regions in a single virialized galaxy,
(3) cooling gas accreting onto a massive dark halo associated
    with an initial onset of starburst at the halo center,
(4) merging bright LAEs with clouds of {\sc Hii} regions,
and
(5) outflowing gas excited by shocks or UV radiation from starbursts and/or mergers.
There is a chance of (1), although no positive evidence of an AGN is found. 
Our object shows the lack of {\sc Nv} line, no line broadening, and no detections 
in X-ray, MIR, submm, and radio bands (Figure \ref{fig:z7blob_multibandsed}). But,
these constraints are not strong enough to
discard the possibility of a hidden AGN.
The case of (2) seems surprising, because it means that such a
large galaxy exists in a very early epoch of $z=6.6$. This galaxy
would have a size of $\gtrsim 17$ kpc and possibly two large star-forming regions 
in a disk seen in our $z'$ band image (\S \ref{sec:gravitational_lensing}). 
However, there is a chance to explain this large galaxy 
in the framework of Cold Dark Matter (CDM) models. 
We estimate properties of the most massive dark halo 
whose number density is the same as our object at $z=6.6$.
Based on the analytic CDM model of \citet{sheth1999},
we find that the dark halo has a radius of 47 kpc and
a mass of $1\times 10^{12} M_\odot$
with a circular velocity of 380 km s$^{-1}$.
All of these values of the dark halo are significantly larger 
than those of our giant LAE measured via radiation, 
indicating that the single-galaxy picture
can be compatible with the CDM model
in terms of halo properties.
But it is not obvious that such a big virialized 
baryonic system at $z=6.6$ can be reproduced in the scheme of CDM model.
The case of (3) is possible with the potentially 
large $EW_0$ of $57-402$\AA. 
Moreover, our DEIMOS spectrum presents a possible weak red peak
at $\sim 9245$\AA\ (Figure \ref{fig:blob_2dspec_1dspec}) which is 
redshifted from the main Ly$\alpha$ peak ($9233$\AA) by $\sim 400$ km s$^{-1}$.
This possible red peak may be similar to the one produced
by IGM infalling into a collapsing cloud, which is claimed
by \citet{dijkstra2006a,dijkstra2006b}. 
On the other hand, the comparable SFR values 
from Ly$\alpha$ and UV continuum (\S \ref{sec:stellar_population})
may not prefer this scenario, 
since the Ly$\alpha$ luminosity
could be explained solely by normal star-forming activities 
that the UV continuum indicates.
However, this argument is not strong due to the underestimation of 
SFR from Ly$\alpha$ luminosity. 
In fact, 
the intrinsic Ly$\alpha$ luminosity could be brighter by a factor of 20
if our giant LAE has diffuse Ly$\alpha$ components similar to
\citeauthor{steidel2000}'s blob 1 (\S \ref{sec:comparisons}).
The explanation of (4) would be reasonable, given 
the possible existence of two peaks in $z'$ band 
with a separation of $1''.1$ (6.0 kpc).
The $NB921$-image profile can be reproduced by
two exponential disks with a reasonably small half-light radius of
$R_{\rm hl}=0''.3$ corresponding to $R_{\rm hl}=1.6$ kpc 
(\S \ref{sec:gravitational_lensing}; cf. 
\citealt{simard1999}).
The merger would induce star-formation activities,
and could produce the bright Ly$\alpha$-line and UV-continuum emission.
Figure \ref{fig:z7blob_specdiff} shows that velocity widths
at $position$ 1 and 2 are similar, implying
that dynamical masses of these components would be comparable.
We may be witnessing the site of a major merger 
near the reionization epoch.
On the other hand, our object shows a Ly$\alpha$ nebula 
potentially larger than the 
isophotal scale with the major axis of $3''.1$ (17 kpc; 
\S \ref{sec:photometric_identification}).
If the Ly$\alpha$ nebula really extends beyond
the $2\sigma$-level isophotal area, it becomes difficult
to explain the Ly$\alpha$ morphology with a profile of two merging LAEs.
The (5) case seems plausible, since our object has
a relatively high SFR and a large stellar-mass
with possible multiple components in $z'$ band (\S \ref{sec:gravitational_lensing}).
In either case of (4) or (5),
our object would be a massive galaxy in formation
with significant star-formation contributing 
to cosmic reionization (cf. \citealt{iliev2006}) 
and/or 
with outflows for the metal enrichment of IGM \citep{bouche2007}.
Since our object has the small velocity offset ($\Delta v=60$km s$^{-1}$)
and the line width ($v_{\rm FWHM}=251$ km s$^{-1}$), the dynamics of 
merger or outflow would have to be well collimated to the direction
perpendicular to the line of sight.

The angular size of the Ly$\alpha$ nebula
is $\gtrsim 17$ proper kpc (\S \ref{sec:gravitational_lensing}),
which is comparable to the diameter of the stellar disk
of the present-day Milky Way.
It is impressive, if we consider that the age of the Universe at $z=6.595$ 
is only 6\% of the one of the present-day Universe. 
Moreover, such an extended Ly$\alpha$ source is very rare in the cosmological volume
only with the number density of $1.2\times 10^{-6}$ comoving Mpc$^{-3}$ at $z=6.6$
(\S \ref{sec:spectroscopic_confirmation}).
If our selection of large Ly$\alpha$ nebula does not miss a significant fraction of 
massive galaxies at this early epoch ($z=6.6$),
our object could be an ancestor of a bright-cluster or cD galaxy,
and should be a good laboratory of massive-galaxy formation
near the reionization epoch.

\subsection{Future Prospects}
\label{sec:future}

In section 4.3, we have found that the currently available data do not 
provide a clear answer to the question about the nature of this object. 
It is obvious that deeper NIR and infrared images of this object 
can be taken with Hubble and Spitzer Space Telescopes 
to constrain SFR, dust extinction, and stellar age, 
which will trace back through the star-formation history of 
this object (e.g. \citealt{yan2006,eyles2007}).
These deep images are critically important to give stronger constraints
on the SED of our object (Figure \ref{fig:z7blob_sed}) that
characterizes SFR and dust extinction,
and to identify Ly$\alpha$ photons 
that are not originated from star-formation activities 
but from others, such as cold accretion.
The high-resolution NIR image of Hubble Space Telescope
is useful to investigate the possibilities of mergers
and outflows. A $J$-band spectrum will test 
the existence of AGN with a relatively strong {\sc Civ} emission line.
We estimate an expected {\sc Civ} flux to be 
$\sim 2$ and $4 \times 10^{-17}$ erg s$^{-1}$ cm$^{-2}$ 
for Seyfert II and QSO, assuming the ratios of 
$f_{\rm CIV}/f_{\rm Ly\alpha}\sim 0.2$,
and
$f_{\rm CIV}/f_{\rm Ly\alpha}\sim 0.5$,
respectively \citep{mccarthy1993}.
These flux limits are achieved by deep NIR spectroscopy
with 8m-class telescopes.
Moreover, we can characterize star-formation
activities and metal enrichment in our Ly$\alpha$ nebula 
by deep submillimeter and millimeter observations
with Atacama Large Millimeter/Submillimeter Array (ALMA)
which will start the operation in 2012 preceded by the early-science operation.
ALMA observations will allow us to investigate emission from dust and molecular-clouds
in our object.
A detection of spatially-extended metal line
from the Ly$\alpha$ nebula could reject the possibility of 
cooling accretion of primordial gas, and provide an independent probe of 
star-formation properties.

We estimate an expected intensity of dust emission in $850\mu$m, $S(850\mu{\rm m})$,
and a flux of molecular CO(6-5) line, $S({\rm CO})$,
assuming typical parameters of local starbursts. 
We start the calculations from our lower-limit of SFR, $34 M_\odot$yr$^{-1}$.
The far-infrared luminosity, $L({\rm FIR})$, 
is calculated from
$SFR [M_\odot {\rm yr}^{-1}]=1.7\times 10^{-10} L({\rm FIR}) [L_\odot]$
\citep{kennicutt1998}.
We obtain $L({\rm FIR})=2.0\times 10^{11} L_\odot$.
Assuming the modified blackbody radiation with
a dust emissivity index of $\beta=1.3$ and 
a dust temperature of $T_{\rm dust}=35.6$ K
\citep{dunne2000},
we estimate the $850\mu$m dust emission from $L({\rm FIR})$
to be $S(850\mu{\rm m})=0.28$ mJy (see, e.g., \citealt{ouchi1999}).
This moderately bright $850\mu$m emission is expected
because of the negative k-correction \citep{blain2002}.
Again from the $L({\rm FIR})$ value with the same modified blackbody radiation,
the dust mass is $M_{\rm dust}=6.0\times 10^{7} M_\odot$
via the relation presented in \citet{debreuck2003}.
The mass of molecular hydrogen is $M(H_2)=3.0\times 10^{9} M_\odot$ which is
calculated with the relation of $M({\rm H_2})/M_{\rm dust}=50$ 
($\sim 25-75$; \citealt{seaquist2004,young1996}).
Finally, the flux of molecular CO(6-5) line is 
$S({\rm CO})=0.1$ Jy km s$^{-1}$
with the assumptions of 
an H$_2$-to-CO conversion factor of 0.8 \citep{downes1998},
and a line ratio of ${\rm CO}(6-5)/{\rm CO}(1-0)=0.5$ \citep{bayet2006}.
In summary, we expect $S(850\mu{\rm m})\gtrsim 0.28$ mJy and 
$S({\rm CO})\gtrsim 0.1$ Jy km s$^{-1}$, considering that 
our SFR is the lower limit ($>34 M_\odot$yr$^{-1}$).
We use ALMA Sensitivity Calculator
\footnote{
http://www.eso.org/sci/facilities/alma/observing/tools/etc/index.html
},
and estimate the on-source integration time
to be $\lesssim 0.2$ and $\lesssim 3$ hours for 
$5\sigma$ detections of a $850\mu{\rm m}$ continuum 
and a CO(6-5) line, respectively. 
Here we assume the large beam size of $1''$ 
in a compact configuration of 50 12m-arrays for the point-source detection,
the band width of 16 GHz, and the CO-line width of $250$ km s$^{-1}$ 
\citep{nishiyama2001} with a 50 km s$^{-1}$ spectral resolution.
Either of an $850\mu$m-thermal continuum or a molecular-CO(6-5) line
may be detected in reasonable observing time
under the assumptions of local starbursts.
If our object does not have dust or molecular gas 
as much as the local starbursts, 
a deficit of dust or molecular-line 
emission would be identified by ALMA observations.
In either case, dust and molecular-gas properties of our object
could be characterized in a few years.

\acknowledgments
We are grateful to 
Robert Antonucci,
George Becker,
Arjun Dey,
Richard Ellis,
Joseph Hennawi,
Masakazu Kobayashi,
Juna Kollmeier,
Ivo Labb{\'e},
Janice Lee, 
Yuichi Matsuda,
Kazuaki Ota, and
Ann Zabludoff
for their useful comments and discussions.
We acknowledge Yuichi Matsuda for providing us the cutout narrow-band 
images of their Ly$\alpha$ blobs.
We thank Michael Cooper who gave us helpful advices on 
the installation and the parameter choice of the spec2d pipeline.
M.O. has been supported via Carnegie Fellowship.
J.S.D. and R.J.M. acknowledge the support of the Royal Society.
C.L.M. thanks the Packard Foundation for their financial support.
This work is based in part on observations made with the Spitzer 
Space Telescope, which is operated by the Jet Propulsion Laboratory, 
California Institute of Technology under a contract with NASA.
Support for this work was provided by NASA through an award issued 
by JPL/Caltech.
This work is supported in part by Department of Energy contract
DE-AC02-76SF00515.
The authors wish to recognize and acknowledge the very significant cultural role 
and reverence that the summit of Mauna Kea has always had within the 
indigenous Hawaiian community.  We are most fortunate to have the opportunity 
to conduct observations from this mountain.

{\it Facilities:} \facility{Subaru (Suprime-Cam)}, \facility{Keck:II (DEIMOS)}, \facility{Magellan:Baade (IMACS)}, \facility{Spitzer (IRAC,MIPS)},



\end{document}